\documentclass[11pt,letterpaper]{article}

\usepackage{amsmath}
\usepackage{array}
\usepackage{dcolumn}
\usepackage{graphicx}
\usepackage{wrapfig}

\newcommand{\br}{\ensuremath{\mathcal{B}}}

\newcommand{\bmes}{\ensuremath{\mathrm{B}}}
\newcommand{\bplus}{\ensuremath{\mathrm{B}^+}}
\newcommand{\bmin}{\ensuremath{\mathrm{B}^-}}
\newcommand{\bz}{\ensuremath{\mathrm{B}^0}}
\newcommand{\dmes}{\ensuremath{\mathrm{D}}}
\newcommand{\dstar}{\ensuremath{\mathrm{D}^\ast}}
\newcommand{\kmes}{\ensuremath{\mathrm{K}}}
\newcommand{\ks}{\ensuremath{\mathrm{K}_S^0}}

\newcommand{\dcalc}[1]{\ensuremath{\mathrm{d}#1}}

\newcommand{\boldx}{\ensuremath{\mathbf{x}}}
\newcommand{\boldt}{\ensuremath{\boldsymbol{\theta}}}

\begin{document}

\pagestyle{empty}
\begin{center}

  {\LARGE
   Neyman \& Feldman-Cousins intervals for a simple problem
   with an unphysical region, and an analytic solution       \\
   \vspace*{2ex}}

  {\large 
   B.D.~Yabsley                            \\[1ex]
   \verb+http://belle.kek.jp/~yabsley/+    \\[1ex]
   \itshape
   Falkiner High Energy Physics Department \\
   School of Physics, University of Sydney \\
   NSW 2006 AUSTRALIA}

  \begin{abstract}
    \noindent
    The new Belle $\phi_3/\gamma$ measurement \texttt{arXiv:hep-ex/0604054},
    based on Dalitz analysis of $\dmes \to \ks\pi^+\pi^-$ 
    in $\bmes^{\pm} \to \dmes^{(*)} \kmes^{(*)\pm}$ decays, 
    uses likelihood ratio ordering to set confidence intervals in $\phi_3$ and the
    $r,\delta$ parameters. This is different to the choice made by BaBar in
    PRL \textbf{95}, 121802 (2005) and \texttt{arXiv:hep-ex/0507101}, and requires 
    additional computation. This Note explains Belle's choice 
    using a related but simpler example: the averaging of two numbers. 
    We find that intervals calculated with likelihood ratio ordering 
    reproduce the analytic solution to this problem, whereas intervals calculated by 
    ordering according to the p.d.f.\ (so-called \emph{Neyman} intervals) do not,
    and show a pathology which is important in our case.

    This document is adapted from a Belle Internal Note.
  \end{abstract}
\end{center}

\setcounter{page}{0}

\clearpage

\pagestyle{plain}

\begin{flushright}
  \small 
  {[Belle-internal labels went here]} \\
  Bruce Yabsley 2006/03/21; edited for \texttt{hep-ex} 2006/04/26
\end{flushright}

\section*{Neyman \& Feldman-Cousins intervals for a simple problem
	with an unphysical region, and an analytic solution}

\subsection*{Introduction}

A known pathology of frequentist methods is that they can give empty confidence intervals
in some cases; more generally, they have trouble handling measurements
with unphysical regions, such as $\sqrt{A^2 + S^2} > 1$ in the analysis of
time-dependent $\bz\to\pi^+\pi^-$ decays.
The so-called Feldman-Cousins approach was introduced (in part) to solve this problem
from first principles.

The following is an illustration of how this works, using a simple example of a measurement
with an unphysical region. The example is inspired by the problem of extracting
the parameters $(\phi_3,\, r,\, \delta)$ from measurements $(x_+,\, y_+,\, x_-,\, y_-)$
in the $\bmes^\pm \to \dmes^{(*)} \kmes^{(*)\pm}$ Dalitz analysis, and was chosen such that
an ``obvious'' right answer already exists, in analytic form.

\subsection*{The problem}

Consider a continuous quantity $\mu$, of which we make two independent measurements
$a$ and $b$. Suppose that each measurement is unbiased and distributed as a Gaussian
with known standard deviation $\sigma = 1$,
so that the probability density of the pair $(a,b)$ is
\begin{equation}
  f(a,b;\,\mu) = \mathcal{G}(a;\, \mu,\, \sigma=1) \cdot \mathcal{G}(b;\, \mu,\, \sigma=1),
  \label{eq-pdf}
\end{equation}
where $\mathcal{G}(x;\,m,\,s) \equiv \frac{1}{s\sqrt{2\pi}}
                                     \exp\left(\frac{-(x-m)^2}{2s^2}\right)$.
Given a measurement $(a,b) = (a_0, b_0)$, what is the confidence interval in $\mu$
for a given confidence level?

\noindent
The obvious solution to this is that the best estimate of $\mu$,
and the corresponding confidence intervals, are given by the simple mean
and the standard error of the mean,
\begin{equation}
  \mu_{\text{est}} = \frac{1}{2}(a_0 + b_0),\;\;
  \sigma_M = \frac{\sigma}{\sqrt{2}}:
  \label{eq-muest-sigmam}
\end{equation}
\begin{itemize}
  \item	the 68.3\% ($1\sigma$) interval should be
	$[\mu_{est}-\sigma_M,\, \mu_{est}+\sigma_M]$,
  \item	the 95.4\% ($2\sigma$) interval should be
	$[\mu_{est}-2\sigma_M,\, \mu_{est}+2\sigma_M]$,
\end{itemize}
and so on. Supposing that we're perverse enough to throw the full frequentist machinery
at the problem, we want to see if we can recover this solution.

In the next section we briefly revisit the frequentist construction of confidence intervals:
if you're already familiar with this (or if such details bore you), 
please skip to page~\pageref{sec-neyman},
where we treat the default or ``Neyman'' implementation, 
followed by Feldman and Cousins' likelihood ratio ordering on page~\pageref{sec-feldman}.
A summary comparison, 
and the application to the $\bmes^\pm \to \dmes^{(*)} \kmes^{(*)\pm}$ Dalitz analysis,
can be found on page~\pageref{sec-comparison}.

\clearpage

\subsection*{Frequentist confidence intervals (for revision: skip this if desired)}
\label{sec-revision}

\begin{wrapfigure}{r}{0.5\textwidth}
  \includegraphics[width=0.48\textwidth]{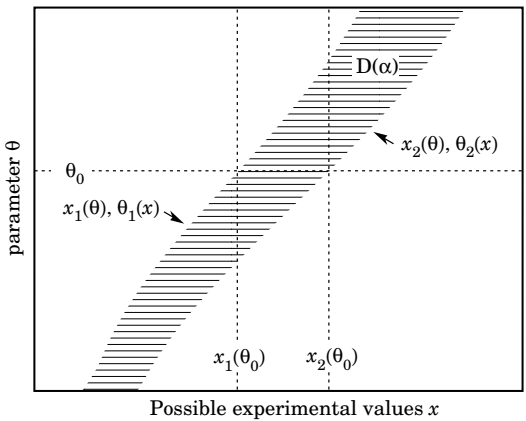}
  \caption{Construction of confidence intervals
	by forming the \emph{confidence belt} $D(\alpha)$.
	From the PDG2004 review [Fig.~32.3].
	This plot assumes monotonic $x_{1,2}(\theta)$,
	defining corresponding functions $\theta_{1,2}(x)$.
}
  \label{fig-confbelt}
\end{wrapfigure}

If $f(x;\,\theta)$ is the p.d.f.\ for a measurement $x$ given a parameter $\theta$,
then for a given confidence level $(1-\alpha)$ we seek values $x_1$, $x_2$ such that
\begin{equation}
  P(x_1 < x < x_2;\, \theta) = 1 - \alpha = \int_{x_1}^{x_2} \dcalc{x} f(x;\,\theta),
  \label{eq-confbelt}
\end{equation}
\emph{i.e.}\ we want a (small) probability $\alpha$ for the measured value $x$ to lie
outside the interval $[x_1,x_2]$. 
In principle this should be done for all possible $\theta$, 
defining functions $x_1(\theta)$, $x_2(\theta)$.
The result is a belt $D(\alpha)$ in $(x,\theta)$,
as shown in figure~\ref{fig-confbelt}.
For a given measurement $x_0$ one draws a vertical line $x=x_0$:
its intersection with $D(\alpha)$ is the confidence interval in $\theta$,
$[\theta_2(x_0),\theta_1(x_0)]$. 
The method is general and can be applied to multidimensional parameters \boldt\
and data \boldx: the integral in Eq.~(\ref{eq-confbelt}) 
is then performed over a region in \boldx.

If the experiment is repeated $n$ times, the measurements $x_0$ and
confidence intervals $[\theta_2(x_0),\theta_1(x_0)]$ will vary, 
but the true value $\theta$ should lie inside the interval 
in a fraction $(1-\alpha)$ of cases. 
The ideal, where this holds for any $\theta$
(in the limit where $n \to \infty$), is called \emph{coverage}:
\begin{equation}
  1-\alpha = P[\theta_2(x;\,\alpha) < \theta <\theta_1(x;\,\alpha)]
  \label{eq-coverage}
\end{equation}
For continuous $f$ this follows from eq.~(\ref{eq-confbelt});
for discrete $f$, or for cases where an approximate method is used,
eq.~(\ref{eq-coverage}) will not hold in general.
If the fraction is smaller than $(1-\alpha)$ for some value $\theta$ then 
the interval-setting method is said to \emph{undercover} for that value, which is bad;
if larger than $(1-\alpha)$, then it \emph{overcovers}, which is conservative
(implying some corresponding loss of power for the method).

Equation~(\ref{eq-confbelt}) is not sufficient to define $D(\alpha)$,
since in general, given $\theta$ there is more than one choice for $(x_1,x_2)$:
infinitely many, if $f$ is continuous. Typically some algorithm is
used to determine $x_{1,2}(\theta)$, 
and hence the confidence interval $[\theta_2(x_0),\theta_1(x_0)]$,
for any measurement $x=x_0$. 
For the case shown in figure~\ref{fig-confbelt} the special choice
\begin{description}
  \item[$x_2 \to +\infty$] always produces an \emph{upper limit}, a special interval with 
                           only an upper edge in $\theta$, since for any measurement $x_0$
			   the interval will include values $\theta\to-\infty$,
			   or to whatever the minimum defined value might be: for example,
			   the ``90\% C.L.\ upper limit $\br_{90}$''
			   on a branching fraction \br\ is the confidence interval
			   $\br \in [0,\br_{90}]$;
  \item[$x_1 \to -\infty$] always produces a \emph{lower limit}, 
                           an interval with only a lower edge in $\theta$, 
			   which will include  $\theta\to+\infty$ (or $1$, or whatever).
\end{description}
In more general cases, confidence intervals need not be simply connected 
(\emph{i.e.}\ they can have gaps, 
contrary to the assumption used to draw figure~\ref{fig-confbelt}); 
and in pathological cases they can be empty (because $x=x_0$ never intersects $D(\alpha)$). 
As we'll see, empty intervals can occur even for the very simple case considered in this note.

\clearpage

\subsection*{Intervals with ``Neyman'' ordering}
\label{sec-neyman}

\begin{figure}[!b]
  \begin{center}
    \includegraphics[width=0.85\textwidth]{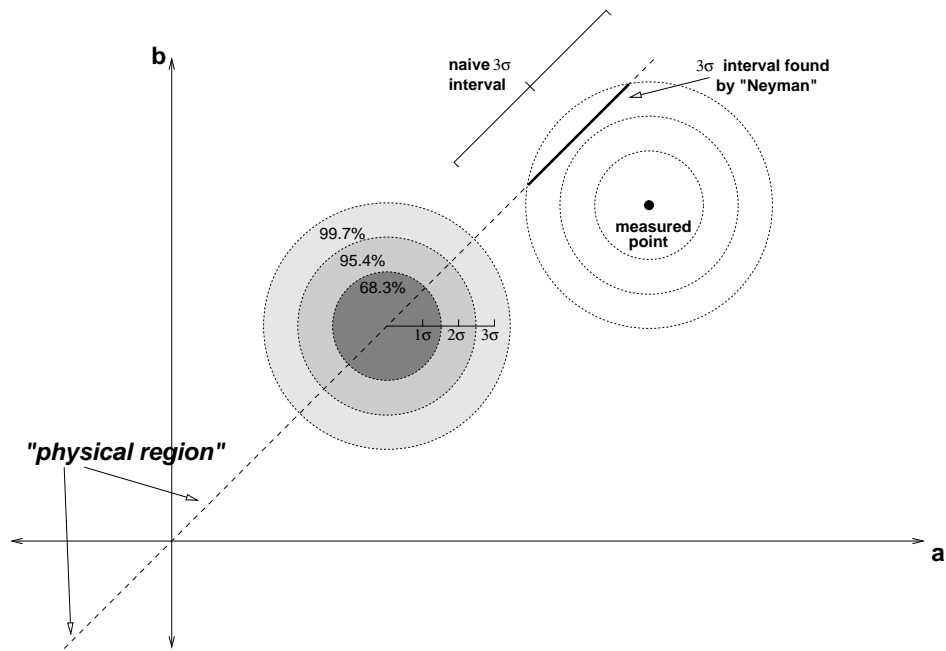}
  \end{center}
  \caption{Construction of confidence intervals using ``Neyman'' ordering.}
  \label{fig-neyman}
\end{figure}

So, for any given parameter $\theta$ we must integrate the p.d.f.\ $f(x;\,\theta)$
for the measurement $x$ until we have an area equal to the confidence level $(1-\alpha)$
(see eq.~(\ref{eq-confbelt})).
A straightforward way to do this is to set the probabilities at the boundaries 
$x_1$ and $x_2$ to be the same, $P(x_1) = P(x_2)$.
For a symmetric function $f(x;\,\theta)$ 
this is gives a \emph{central interval}
\begin{equation}
  P(x < x_1;\, \theta) = \int_{-\infty}^{x_1} \dcalc{x} f(x;\,\theta) = \frac{\alpha}{2}
                       = \int_{x_2}^{\infty}  \dcalc{x} f(x;\,\theta) =
  P(x_2 < x;\, \theta),
  \label{eq-neyman-symmetric}
\end{equation}
with equal area in each tail. For more general functions $f$ or for functions in $n>1$
dimensions, we choose a domain in \boldx\ beginning with the points of highest probability,
and then including lower-probability points in turn, until the desired area is achieved,
\begin{equation}
  \int\limits_{f>f_{\text{min}}(\alpha)} \dcalc{\boldx} f(\boldx;\,\boldt) = 1 - \alpha. 
  \label{eq-neyman-general}
\end{equation}
This simplest kind of \emph{ordering} of the points included in the integral 
is called by many people \emph{Neyman} ordering, although as far as I know Neyman is 
responsible for the general method described in the previous section, not for the choice
(\ref{eq-neyman-general}).

Applying this to our $f(a,b;\,\mu)$ of eq.~(\ref{eq-pdf}), 
we have the integral of a 2D Gaussian,
\begin{equation}
  1 - \alpha   = \iint\limits_{f>f_{min}(\alpha)} \dcalc{a}\,\dcalc{b} \frac{1}{2\pi} 
                 \exp \left(
		   \frac{-\left[ (a-\mu)^2 + (b-\mu)^2 \right]}{2} 
		 \right),
  \label{eq-neyman-fmin}
\end{equation}
which we solve as usual substituting $(a-\mu)=\rho\cos\psi$, $(b-\mu)=\rho\sin\psi$:

\small
\begin{align}
  1 - \alpha   = \iint\limits_{f>f_{\text{min}}(\alpha)} \dcalc{\rho}\,\dcalc{\psi}\,
                 \left| J(\rho,\psi) \right|\,\frac{1}{2\pi} 
                 \exp \left( \frac{-\rho^2}{2} \right)                     
 	     & = \int\dcalc{\psi}
		 \int\limits_{\rho<\rho_{\text{max}}(\alpha)}\dcalc{\rho}\,
		 \rho\frac{1}{2\pi} 
                 \exp \left( \frac{-\rho^2}{2} \right),          \nonumber \\
	     & = \left[ 
		   -\exp \left( \frac{-\rho^2}{2} \right) 
		 \right]_0^{\rho_{\text{max}}(\alpha)}           \nonumber \\
  \rho_{\text{max}}(\alpha)
             & = \sqrt{-2\ln\alpha}.
  \label{eq-neyman-rmax}
\end{align}
\normalsize
For any true value $\theta=\mu$, the confidence belt $D(\alpha)$ includes 
all points in the measurement space $(a,b)$ within a circle of radius $\rho_{\text{max}}$
around $(\mu,\mu)$,
\begin{center}
  \begin{tabular}{lclrr}
    $\alpha$   &  1D equivalent  &  $\rho_{\text{max}}$  &  $\rho_{\text{max}}^2$ 
                                                                        & \multicolumn{1}{l}{$l$ (see below)} \\ \hline
    $0.3173$   &  ``$1\sigma$''  &  $1.515$              &  $2.296$     & 1.138           \\
    $0.1$      &  90\% CL        &  $2.146$              &  $4.605$     & 1.384           \\
    $0.05$     &  95\% CL        &  $2.448$              &  $5.991$     & 1.466           \\
    $0.0455$   &  ``$2\sigma$''  &  $2.486$              &  $6.180$     & 1.476           \\
    $0.0027$   &  ``$3\sigma$''  &  $3.439$              & $11.829$     & 1.682           \\ \hline
  \end{tabular}
\end{center}
where the $\rho_{\text{max}}^2$ correspond to the $\Delta \chi^2$
values for \emph{two}-dimensional confidence intervals.  For what we
think of as an $n\sigma$ interval, $\rho_{\text{max}} > n\sigma$
as shown in the lower-left of Figure~\ref{fig-neyman}.

A measured point $(a_0,b_0)$ will be inside the confidence belt 
$D(\alpha)$ for values $\theta = \mu$ where the distance  
$\sqrt{(a_0-\mu)^2 + (b_0-\mu)^2} \leq \rho_{\text{max}}$.
We can therefore find the confidence interval in $\mu$ by taking the intersection
of the $a=b$ line with a disk of radius $\rho_{\text{max}}$ drawn around the 
measurement $(a_0,b_0)$. Fig.~\ref{fig-neyman} showns an example
where the measurement is $3\sigma$ away from the physical line: 
the $68.3\%$ and $95.4\%$ disks do not intersect $a=b$,
so the corresponding confidence intervals are \emph{empty}. 
This reflects the low probability 
for the measurement to be this far away from the $a=b$ line.

As for the $99.7\%$ interval, our analytic solution 
(Eq.~(\ref{eq-muest-sigmam})), tells us that
it should be centered at $\mu_{\text{est}} = \frac{1}{2}(a_0 + b_0)$,
with a half-width $3\sigma_M = \frac{3}{\sqrt{2}}\sigma = \frac{3}{\sqrt{2}}$. 
The corresponding $a=b$ line segment runs
  from $(\mu_{\text{est}}-{3}/{\sqrt{2}},\mu_{\text{est}}-{3}/{\sqrt{2}})$
  to   $(\mu_{\text{est}}+{3}/{\sqrt{2}},\mu_{\text{est}}+{3}/{\sqrt{2}})$:
$2\times3=6$ units long.
The Fig.~\ref{fig-neyman} construction gives 
the correct central point $(\mu_{\text{est}},\mu_{\text{est}})$,
but a half-width on the plane of $\sqrt{\rho_{\text{max}}^2-3^2} = 1.682$,
shorter than the correct value of 3.
(The half-width in $\mu$ is smaller by a factor of $\sqrt{2}$, 
which can be [to me] confusing.)

The use of so-called Neyman ordering can thus give us empty confidence
intervals, or intervals that are too narrow, if the measured values
$a_0$ and $b_0$ are far apart.  If, on the other hand, the measured
values are close to each other, the interval can be too \emph{wide}:
for an ideal measurement $a_0=b_0=\mu$, the intersection on
Figure~\ref{fig-neyman} would be $2\rho_{\text{max}} = 6.878$ units
long for a 99.7\% confidence interval, to be compared with the correct
value of 6.0. An $n\sigma$ interval (with C.L.\ $\alpha_n$) will have 
the correct width for data $l\sigma$ away from the physical line
$a=b$, where $l=\sqrt{-2\ln(\alpha_n) - n^2}$: 
this gives a distance $l=1.68\sigma$ for a $3\sigma$ interval. 
Other values are included in the table above.

On average, the true value will in fact lie within the confidence
interval 99.7\%, 95.4\%, 68.3\% (or whatever) of the time, as it must
by construction.  But in some cases, where the interval is empty, we
\emph{know} that the true value is not inside the interval,
and so the exercise has turned out to be useless for our purposes.
Therefore, while the construction is ``correct'' in a technical sense, 
it is clearly pathological.


\clearpage

\subsection*{Intervals with likelihood-ratio ordering (a.k.a.\ ``Feldman-Cousins'')}
\label{sec-feldman}

A different way of choosing the integration domain in \boldx\ 
($\boldx = (a,b)$, in our example) was advocated in a paper by Feldman and Cousins.
One can actually find the principle written down in an (old) standard reference,
but until recently it does not seem to have been implemented. The argument is as
follows: 

Given a parameter \boldt\, for any point \boldx\ we have a decision to make: 
does this point belong in the confidence belt, or not?
The appropriate way 
to make such a decision is based not on a likelihood but on a likelihood \emph{ratio},
\begin{equation}
  \lambda = \frac{f(\boldx;\;\boldt)}{f(\boldx;\;\boldt_{\text{best}}(\boldx))},
  \label{eq-lrat-general}
\end{equation}
where we compare the likelihood of the parameter \boldt\ given data \boldx,
to the likelihood of the \emph{best possible} parameter for those data,
$\boldt_{\text{best}}(\boldx)$.\footnote{Note the distinction:
  The \emph{likelihood} of the parameter given the data is numerically equal to
  the \emph{probability} of the data, given the appropriate value of the true 
  parameter. In frequentist statistics the true parameter ``is what it is''
  and it is not sensible to assign a probability to it.}  
If $f(\boldx;\;\boldt)$ is small compared to $f(\boldx;\;\boldt_{\text{best}})$,
then the parameter \boldt\ is relatively unlikely, given the data. However if
both $f(\boldx;\;\boldt)$ and $f(\boldx;\;\boldt_{\text{best}})$ are small, 
and their ratio $\lambda \simeq 1$, then the low likelihood for \boldt\ does not
matter: even the best parameter value $\boldt_{\text{best}}$ is unlikely,
so there is no reason to exclude this point from the confidence belt.

Using \emph{likelihood ratio ordering} we choose a domain in \boldx\ beginning
with the points of highest likelihood ratio $\lambda$ from Eq.~(\ref{eq-lrat-general}), 
and then including lower-$\lambda$ points in turn, until the desired area is achieved,
\begin{equation}
  \int\limits_{\lambda>\lambda_{\text{min}}(\alpha)} \dcalc{\boldx} f(\boldx;\,\boldt) = 1 - \alpha. 
  \label{eq-feldman-general}
\end{equation}
In general, equation~(\ref{eq-lrat-general}) requires a minimization step for each \boldx, 
and thus more
computation than ordering by probability \`{a} la ``Neyman''. 
A strategy to minimise computation may be necessary to make the method tractable,
although the difficulty should not be exaggerated: it has been done at Belle in a 
number of seemingly difficult cases.

Our problem is simple enough to handle as a high-school exercise: from Eq.~(\ref{eq-pdf}),
\begin{align}
  f(a,b;\;\mu) & = \frac{1}{2\pi} 
                   \exp \left(
		     -\left[ (a-\mu)^2 + (b-\mu)^2 \right] / 2 
		   \right),                                 \nonumber \\
	       & = \frac{1}{2\pi} 
                   \exp \left(
		      \left[ \left( \{u+v\}/\sqrt{2}-\mu \right)^2 + 
		             \left( \{u-v\}/\sqrt{2}-\mu \right)^2 \right] / 2 
		   \right),                                 \nonumber \\
	       & = \frac{1}{2\pi} 
                   \exp \left(
		     -\left[ (u-\sqrt{2}\cdot\mu)^2 + v^2 \right] / 2 
		   \right),                                 \nonumber \\
 f'(u,v;\;\mu) & = \mathcal{G}(u;\,\sqrt{2}\cdot\mu,\,\sigma=1) \cdot
		   \mathcal{G}(v;\,0,\,\sigma=1)            \label{eq-f-uv}
\intertext{where we use rotated coordinates}
  u            & = \frac{1}{\sqrt{2}} (a+b),\;\;
  v              = \frac{1}{\sqrt{2}} (a-b). \label{eq-uv}
\end{align}
In evaluating Eq.~(\ref{eq-lrat-general}), 
we compare $f'(u,v)$ with the best-fitting value at that point,
\begin{figure}[!b]
  \begin{center}
    \includegraphics[width=0.85\textwidth]{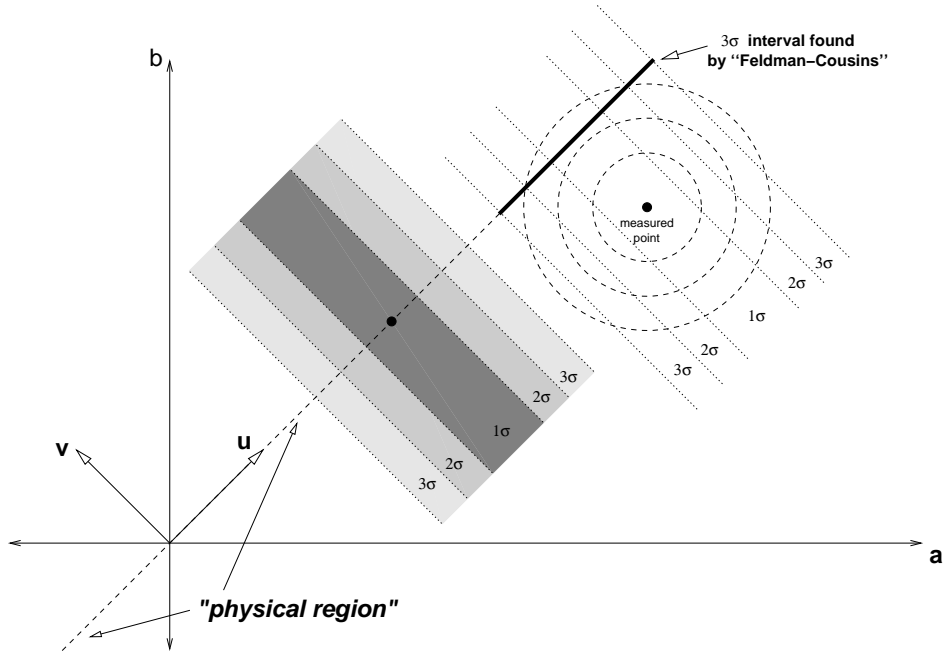}
  \end{center}
  \caption{Construction of confidence intervals using likelihood ratio ordering.}
  \label{fig-feldman}
\end{figure}
\begin{align}
  f^\prime_{\text{best}}(u,v) & = \mathcal{G}(u;\,\sqrt{2}\cdot\mu_{\text{best}},\,1) \cdot
		                  \mathcal{G}(v;\,0,\,1) \nonumber \\
                              & = \mathcal{G}(u;\,u,\,1) \cdot
		                  \mathcal{G}(v;\,0,\,1) \nonumber \\
                              & = 1 \cdot
		                  \mathcal{G}(v;\,0,\,1),
  \label{eq-fbest} \\
\intertext{since the best-fitting estimate of the parameter at $(u,v)\equiv(a,b)$ is}
  \mu_{\text{best}}           & = \mu_{\text{est}} = \frac{1}{2}(a+b) = \frac{1}{\sqrt{2}} u,
  \label{eq-mubest-u}
\end{align}
where the Gaussian reaches its peak. The likelihood ratio is thus
\begin{equation}
  \lambda = \frac{\mathcal{G}(u;\,\sqrt{2}\cdot\mu,\,1) \cdot \mathcal{G}(v;\,0,\,1)}
                 {1                                     \cdot \mathcal{G}(v;\,0,\,1)}
          = \mathcal{G}(u;\,\sqrt{2}\cdot\mu,\,1) 
   \label{eq-lrat-f}
\end{equation}
and the orthogonal coordinate $v$ drops out of the problem.
For any true parameter $\theta=\mu$, then, the confidence belt is given by 
the \emph{one-dimensional} integral
\begin{equation}
  \int\limits_{u=\sqrt{2}\,(\mu-\delta)}^{u=\sqrt{2}\,(\mu+\delta)} \dcalc{u}\; 
    \mathcal{G}(u;\,\sqrt{2}\cdot\mu,\,1) 
  = 1 - \alpha; 
  \label{eq-feldman-f}
\end{equation}
following the same procedure as before (see Fig.~\ref{fig-feldman})
we find an interval 
$\mu \in \left[\mu_{est}-\frac{1}{\sqrt{2}},\, \mu_{est}+\frac{1}{\sqrt{2}} \right]$
at the 68.3\% CL, and
$\mu \in \left[\mu_{est}-\frac{2}{\sqrt{2}},\, \mu_{est}+\frac{2}{\sqrt{2}} \right]$
at 95.4\%, and so on, whatever the data: this is the result we wanted.

\clearpage

\subsection*{What just happened? or, Why do the intervals differ?}
\label{sec-comparison}

Comparing the ``Feldman-Cousins'' procedure (figure~\ref{fig-feldman}) 
to the ``Neyman'' case (figure~\ref{fig-neyman}),
we see that the likelihood ratio ordering keeps points close to 
the best estimate of $\mu$, $\mu_{\text{best}} = \frac{1}{2}(a_0+b_0) = \frac{1}{\sqrt{2}}u_0$,
along the $u$-axis,
even if they are far away from the expected value of zero in the orthogonal coordinate
$v = \frac{1}{\sqrt{2}}(a-b)$ \ldots
\emph{since the difference $(a_0 - b_0)$ is irrelevant to the estimation of the underlying parameter $\mu$}. 
Points with large $|v|$ are indeed improbable, 
but this does not help us discriminate between different $\mu$ values: 
only the relative likelihood along the $u$-axis does that.
By ignoring this distinction, 
the apparently natural procedure of ordering-by-$f(a,b;\,\mu)$ 
produces intervals that vary with $|v|$, 
disappearing when it is large.
At the price of computation (in the general case),
ordering by the likelihood ratio in Eq.~(\ref{eq-lrat-general}) 
automatically makes the distinction 
between relevant ($u$) and irrelevant ($v$) information.
From equations~(\ref{eq-f-uv}), (\ref{eq-fbest}), and (\ref{eq-lrat-f}),
it's clear that this will also work when making $3,4,\ldots n$ measurements
according to a single mean $\mu$, 
and it should apply in general when the dimension of the data $\boldx = \{x_i\}$ 
exceeds the dimension of the parameters $\boldt = \{\theta_j\}$.\footnote{
  Strictly speaking, it's only clear that the logic used here will apply
  for cases where the likelihood can be factorised into functions of the 
  ``relevant'' and the ``irrelevant'' coordinates. I'm not sure
  whether a more fundamental argument can be made. For the special case
  where the true parameter $\theta$ can only take on two discrete values,
  the likelihood ratio condition Eq.~(\ref{eq-lrat-general}) gives the best
  possible discrimination between them, according to the Neyman-Pearson
  theorem. For more general cases, the likelihood ratio test is at least
  considered a good criterion to try by default. 
  (See Eadie \emph{et al.}\ from the Extra Reading.)}

Put another way, given that the problem is overconstrained---one parameter $\mu$ 
for two measurements $a$ and $b$---we can consider goodness-of-fit as well as 
parameter estimation, and ask two separate questions:
\begin{enumerate}
  \item Given our hypothesis, what can we say about the parameter $\mu$?
  \item Is our hypothesis, of independent Gaussian measurements
        $\mathcal{G}(a;\,\mu,\,1)$ and $\mathcal{G}(b;\,\mu,\,1)$,
        consistent with the data?
\end{enumerate}
Likelihood ratio ordering concentrates ruthlessly on question 1.
Question 2 is left for us to consider ourselves, as a matter of due
diligence.  Even if there is some suggestion of a poor fit to the data
($|v| \gg 1$), this is not a reason to bollocks up the confidence
intervals in $\mu$, since question 1 is valid on its own terms.\footnote{This
  distinction is discussed in section~IV.C of the Feldman-Cousins paper:
  ``An advantage of our intervals [is that they] effectively decouple
  the confidence level used for a goodness-of-fit test from the
  confidence level used for confidence interval construction \ldots''}

\subsection*{Application to the $\bmes^\pm \to \dmes^{(*)} \kmes^{(*)\pm}$ Dalitz analysis}
\label{sec-application}

This is relatively straightforward.
In Anton's analysis we are interested in finding 
$(\phi_3,\, r,\, \delta)$---especially $\phi_3$, of course---but 
in order to have well-behaved measurements\footnote{For 
  practical reasons we want quantities whose PDF is approximately a 
  Gaussian, with small (if any) bias. The cartesian coordinates meet
  these criteria, whereas the polar coordinates (esp.\ $r$) do not.}
we choose to measure
$(x_+,\, y_+)$ from the \bplus\ sample and 
$(x_-,\, y_-)$ from the \bmin\ sample,
where $x_\pm = r_\pm \cos(\pm\phi_3+\delta)$
and   $y_\pm = r_\pm \sin(\pm\phi_3+\delta)$.

This is one measurement too many. If we had infinite precision we would
always find $r_+ = r_-$: in practice we always get $r_+ \neq r_-$, 
and a result that is ``unphysical'' in that sense.\footnote{\emph{Cf.}\ 
  the $\phi_2$ analysis using $\bz\to\pi^+\pi^-$, where the dimensionality
  of the parameters and the data was the same, and we had a physical region
  $\sqrt{A^2+S^2} \leq 1$.}
To get the example in this note, take $\mu=r$, $a=r_+$, $b=r_-$,
forget about $r$ needing to be positive, and drop the other quantities. 
That's it.

I made a brief attempt to extend the analytic study to the full problem,
to understand the effect on Neyman intervals
in  
$\phi_3$ and $\delta$ as one moves towards or away from the physical region 
\ldots and only managed to give myself a headache. Clearly the intervals will be affected, 
and presumably in the same sense as those for $r$ (becoming narrower or wider as appropriate).
But the full problem is substantially more complicated than averaging two numbers. 

As to the actual results:
as you know, Anton uses 
three separate decay modes, so the measurement is done three times over,
with $\phi_3$ in common.
The $\bmes^\pm\to \dmes\kmes^\pm$ result is close to the physical case,
whereas the $\bmes^\pm\to \dstar \kmes^\pm$ and $\dmes\kmes^{\ast\pm}$
results are not.

\subsection*{Conclusion}
\label{sec-conclusion}

We already know how to average two numbers $a$ and $b$.
Since this problem is trivial, 
but also resembles estimating 
$(\phi_3,\, r,\, \delta)$ from measurements $(x_+,\, y_+,\, x_-,\, y_-)$
in one important aspect, it provides a suitable test-case for our
statistical method.

We find that likelihood-ratio ordering gives the correct confidence intervals
for the (unknown true) value $\mu$ which the average is used to estimate.
It does this by building a confidence belt along the $\frac{1}{\sqrt{2}}(a+b)$ axis,
which gives information about $\mu$, 
and automatically ignoring the orthogonal axis which only has information on 
goodness-of-fit.

If instead we order according to the p.d.f.\ (``Neyman'' ordering) both axes are
treated equally. If the measurement is close to the ``physical'' case $a=b$, 
the resulting confidence intervals are too wide; if the measurement is far away,
$|a-b| \gg \sigma$, the intervals are too narrow, and can even become
empty, which is pathological.

\subsection*{Further reading}

\begin{itemize}
  \item A primer on confidence intervals, and statistics generally: \\
        Section 32 of the \emph{Review of Particle Physics},
	S.~Eidelman \emph{et al.} (PDG), \emph{Phys.\ Lett.} \textbf{B 592}, 1 (2004).
	Notation on page~\pageref{sec-revision} is chosen to match that in the reference.

  \item {[some Belle-internal things went here]}

  \item Intervals based on likelihood ratio ordering: \\
        Gary~J.~Feldman, Robert~D.~Cousins, 
	``Unified approach to the classical statistical analysis of small signals'',
	\emph{Phys.\ Rev.}\ \textbf{D 57}, 3873 (1998).

  \item Hypothesis testing, the Neyman-Pearson test, and related issues: \\
        Chapter 10 of ``Statistical methods in experimental physics'', 
	W.T.~Eadie, D.~Drijard, F.E.~James, M.~Roos, B.~Sadoulet
	(Amsterdam: North-Holland, 1971).
	There are doubtless better references, 
	but this book is one of the standard ones used by particle physicists. 

\end{itemize}

\clearpage

\subsection*{About this note}

This note arose out of an intermittent discussion with Tim Gershon and others
during Belle-internal review of Anton Poluektov's 
$\bmes^\pm \to \dmes^{(*)} \kmes^{(*)\pm}$ Dalitz analysis.
An early version was shown during a refereeing meeting in July 2005.
\vspace*{1.0ex}

\noindent
The fact that an unphysical region results when $n$ measurements are used
to estimate $m$ parameters, $m<n$, and the problem this causes for 
Neyman intervals, was noticed and urged by Anton.
It seems a lot more obvious in retrospect than it did at the time.
\vspace*{1.0ex}

\noindent
The averaging example presented here is original, as far as I know.     

\end{document}